\documentclass[a4paper,11pt]{article}
\pdfoutput=1 

\usepackage{jinstpub} 

\usepackage[LGRgreek]{mathastext}

\title{Development of the photo-diode subsystem for the HERD calorimeter double-readout.}

\author[a,b]{O. Adriani,}
\author[c]{M. Antonelli,}
\author[d]{A. Basti,} 
\author[a,b]{E. Berti,}
\author[a,b]{P. Betti,}
\author[d,e]{G. Bigongiari,}
\author[b]{L. Bonechi,}
\author[a,b]{M. Bongi,}
\author[c]{V. Bonvicini,}
\author[b]{S. Bottai,}
\author[d,e]{P. Brogi,}
\author[f]{G. Castellini,}
\author[d,e]{C. Checchia,}
\author[g]{J. Casaus,}
\author[h]{X. Cui,}
\author[h]{Y. Dong,}
\author[a,b]{R. D'Alessandro,}
\author[b]{S. Detti,}
\author[g]{F. Giovacchini,}
\author[i,b]{N. Finetti,}
\author[d,e]{P. Maestro,}
\author[d,e]{P.S. Marrocchesi,}
\author[h]{X. Liu,}
\author[g]{J. Marin,}
\author[g]{G. Martinez,}
\author[b]{N. Mori,}
\author[b,1]{L. Pacini,\note{Corresponding author.}}
\author[b]{P. Papini,}
\author[c]{C. Pizzolotto,}
\author[f,b]{S. Ricciarini,}
\author[a]{P. Spillantini,}
\author[b]{O. Starodubtsev,}
\author[e]{L. Stiaccini,}
\author[h]{Z. Tang,}
\author[b]{A. Tiberio,}
\author[b]{E. Vannuccini,}
\author[g]{M. Velasco,}
\author[h]{R. Wang,}
\author[h]{Z. Wang,}
\author[h]{M. Xu,}
\author[c]{G. Zampa,}
\author[c]{N. Zampa}
\author[h]{and L. Zhang}

\affiliation[a]{Department of Physics and Astronomy, University of Florence, I-50019 Sesto Fiorentino, Florence, Italy}
\affiliation[b]{INFN sezione di Firenze, I-50019 Sesto Fiorentino, Florence, Italy} 
\affiliation[c]{INFN Sezione di Trieste, Padriciano 99, I-34149 Trieste, Italy}
\affiliation[d]{INFN Pisa,	Largo B. Pontecorvo, 3 - 56127 Pisa, Italy}
\affiliation[e]{Department of Physical Sciences, Earth and Environment, University of Siena, I-53100 Siena, Italy}
\affiliation[f]{IFAC (CNR), via Madonna del Piano 10, I-50019 Sesto Fiorentino (Firenze), Italy}
\affiliation[g]{Centro de Investigaciones Energ\'{e}ticas, Medioambientales y Tecnol\'{o}gicas (CIEMAT), E-28040 Madrid, Spain}
\affiliation[h]{Key Laboratory of Particle and Astrophysics, Chinese Academy of Sciences, Beijing, China}
\affiliation [i]{Department of Physical and Chemical Sciences, University of L'Aquila, Via Vetoio, Coppito, 67100 L'Aquila, Italy}

\emailAdd{lorenzo.pacini@fi.infn.it}

\abstract{The measurement of cosmic-ray individual spectra provides unique information regarding the origin and propagation of astro-particles. Due to the limited acceptance of current space experiments, protons and nuclei around the "knee" region ($\sim1\ PeV$) can only be observed by ground based experiments. Thanks to an innovative design, the High Energy cosmic-Radiation Detection (HERD) facility will allow direct observation up to this energy region: the instrument is mainly based on a 3D segmented, isotropic and homogeneous calorimeter which properly measures the energy of particles coming from each direction and it will be made of about 7500 LYSO cubic crystals. The read-out of the scintillation light is done with two independent systems: the first one based on wave-length shifting fibers coupled to Intensified scientific CMOS cameras, the second one is made of two photo-diodes with different active areas connected to a custom front-end electronics. This photo-diode system is designed to achieve a huge dynamic range, larger than $10^7$, while having a small power consumption, few mW per channel. Thanks to a good signal-to-noise ratio, the capability of a proper calibration, by using signals of both non-interacting and showering particles, is also guaranteed. In this paper, the current design and the performance obtained by several tests of the photo-diode read-out system are discussed.}

\arxivnumber{????} 

\begin{document}
\maketitle
\flushbottom

\section{The HERD facility.}
\label{sec:HERD_facility}

\begin{figure}[t]
 	\centering 
 	\includegraphics[width=.41\textwidth]{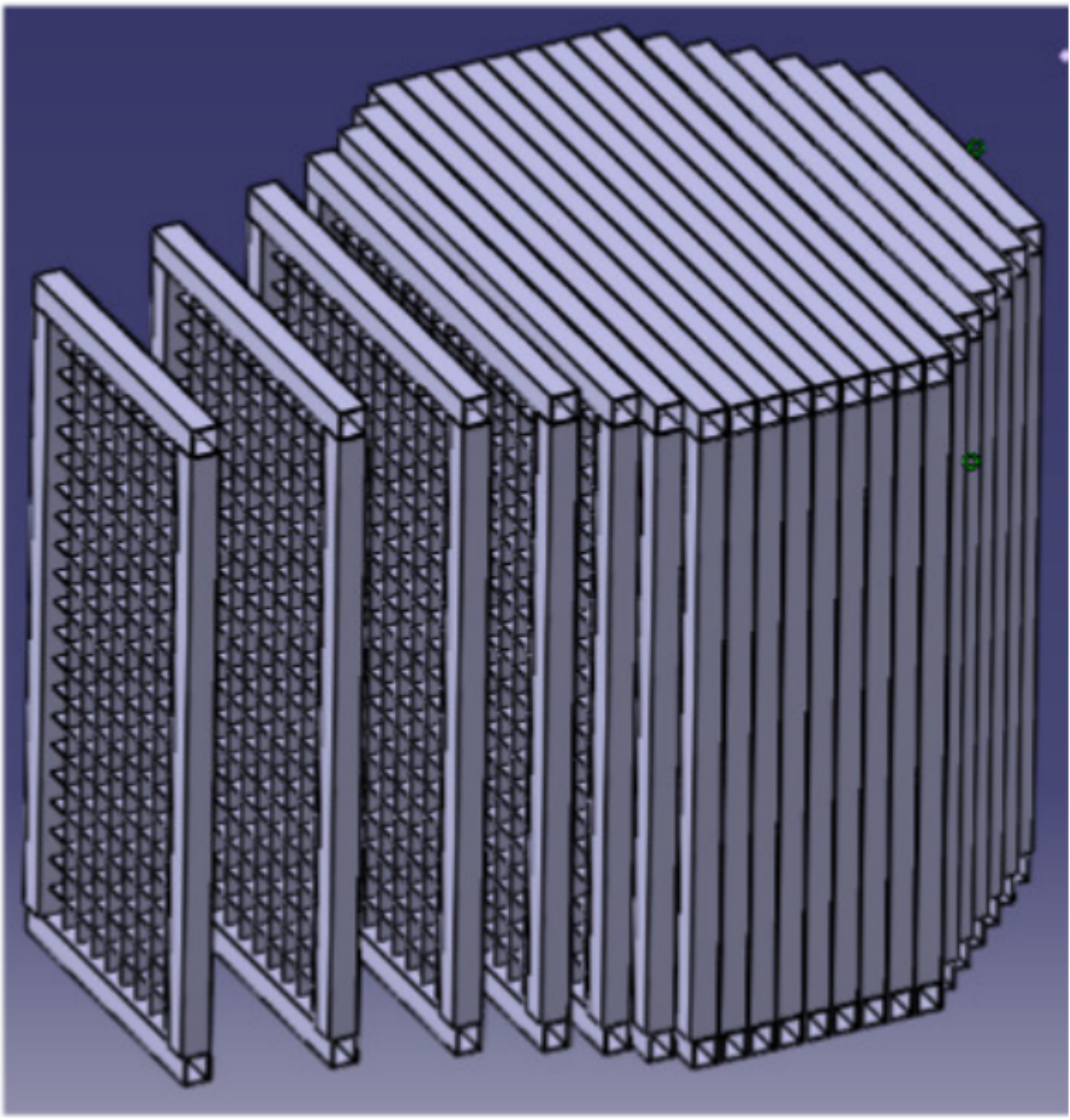}
 	\qquad
 	\includegraphics[width=.36\textwidth]{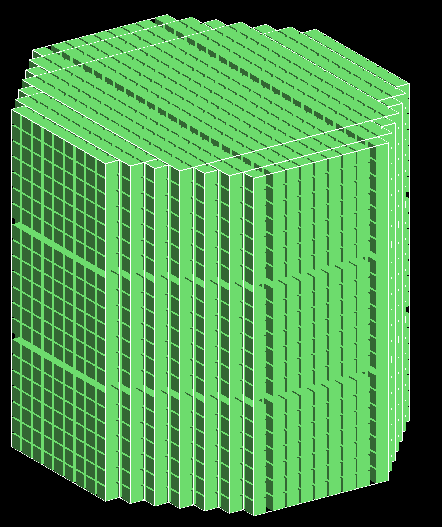}
 	\caption{\label{fig:caloImages} Preliminary arrangement of the CALO layers (left) and image of the LYSO cubes obtained with GEANT4 viewer.}
\end{figure}

The High Energy cosmic-Radiation Detection (HERD)\cite{ICRC2021_Gargano} facility is a space experiment which will be installed aboard the Chinese Space Station (CSS) around 2027. The collaboration includes several Chinese and European institutes and universities. Current cosmic-ray (CR) direct measurements (e.g. \cite{CALET_proton}\cite{DAMPE_proton}) provide information about CR proton spectrum up to $\sim100\ TeV$. Thanks to an innovative design, HERD will be capable to extend direct measurements of CR spectra up to regions not yet explored with space detectors. The experiment will observe protons up the "knee" region (about $\sim 1\ PeV$), where a change of the spectral index of the all-particle spectrum is expected. Furthermore, HERD will accurately measure the individual nuclei spectra up to hundreds of TeV/nucleon: these measurements are very important to enhance our understanding on acceleration and propagation of CRs. HERD will also achieve the first direct observation of the all-electron (electron+positron) flux up to tens of TeV, where new spectral features due to CR nearby sources are expected by several theoretical models (e.g. \cite{HighEnEle_model}), Finally, the experiment will measure the photon spectrum from $\sim500\ MeV$ to $100\ TeV$ and will monitor the gamma-ray sky with a large ``field of view''.\\
CR experiments currently operating in space are telescopes that can properly detect particles entering from the top face. Instead, the innovative design of the HERD experiment allows for the correct reconstruction of particles entering the detector from all sides but the bottom one. The instrument consists of a large acceptance calorimeter and several sub-detectors\cite{ICRC2021_Gargano} which mainly reconstruct the incoming particle charge and primary track. A Transition Radiation Detector (TRD)\cite{TRD} is also present: this will be employed to cross-calibrate the calorimeter energy scale for  $1\ TeV$ protons since the TRD will be calibrated on ground with electron beams with energies around the GeV scale, which correspond to gamma factors of TeV protons.\\
The main HERD detector is a homogeneous, isotropic, 3D segmented calorimeter (CALO), composed of cubic scintillator crystals: as shown by the CaloCube collaboration\cite{Calocube_proc1}\cite{Calocube_proc2}, such a design achieves an acceptance which is about three times larger than a typical calorimeter with the same total mass and volume. Several scintillating crystals were considered for this application: as described in \cite{Calocube_bottai}, LYSO turned out to be one of the best performing scintillating materials for space borne calorimeter. This is because LYSO offers better performances in terms energy resolution and effective geometric factor thanks to its short radiation length ($X_0$) and nuclear interaction length ($\lambda_I$) (e.g. these are similar or shorter than BGO ones), which guarantee a good shower containment. In addition, this crystal has a large light yield (e.g. larger than BGO one) which is beneficial for our application. Table \ref{tab:LYSO} shows important parameters of this scintillator.
\begin{table}[h]
\begin{tabular}{| c | c | c | c | c | c | c |}
	\hline	
	Material & $\rho\ (g/cm^3)$ & $\lambda_I\ (cm)$ & $X_0\ (cm)$ & $\lambda_{max}\ (nm)$ & $\tau\ (ns)$ & L. Y. (photons/keV)\\ \hline	
	LYSO & $7.25$ & $21$ & $1.1$ & $\sim425$ &  $40$ & $\sim30$ \\ \hline	
\end{tabular}
\caption{\label{tab:LYSO} Main parameters of LYSO crystals: $\rho$ is the density, $\lambda_I$ is the nuclear interaction length, $X_0$ is the radiation length, $\lambda_{max}$ is the emission spectrum peak, $\tau)$ is the decay time and L. Y. is the light yield.}
\end{table} \\
This material will be used for about 7500 cubic crystals composing the HERD calorimeter. The cube edge is $3\ cm$, equal to $2.6\ X_0$ and 1.4 Molière radius. As shown in fig. \ref{fig:caloImages}, CALO is made of vertical layers of different sizes, the largest ones containing $21\times21$ crystals. Due to the peculiar layer arrangement, the CALO external envelop is shaped as an octagonal prism: the depth for vertical particles is about $55\ X_0$ and 3 $\lambda_I$. The HERD collaboration is currently working on the CALO geometry optimization, thus the configuration discussed in this paper could be slightly different with respect to the future design for the flight model.
\\
The performance of the CALO\cite{CaloICRC2021} has been studied with GEANT4 Monte Carlo (MC) simulations. The effective geometric factor (GF), i.e. the product of geometric factor and reconstruction efficiency, is better than $2\ m^2sr$ for electrons and $1\ m^2sr$ for protons and nuclei. The corresponding energy resolution at $10\ TeV$ is about $~2\%$, $30\%$ and $22\%$ for electrons, protons and Carbon nuclei, respectively. The electron/proton discrimination is excellent thanks to the 3D sampling of the shower: a residual proton contamination of few percent for high energy electrons is obtained by exploiting the different longitudinal and lateral profile of the hadronic and electromagnetic showers. The MC simulation also shows that a saturation level of at least $\sim200\ TeV$ for the signal of a single crystal is required to properly measure hadronic showers corresponding to few $PeV$ protons: this is one of the main challenges for the design of the CALO read-out system.\\
The LYSO crystal scintillation light is collected with two independent systems. As mentioned in several articles (e.g. \cite{CALET_electron}\cite{DAMPE_electron}), the systematic uncertainty related to the energy scale calibration has sizable impact on the flux measured with calorimetric space experiments. The HERD "double read-out" scheme provides unique capabilities of cross-calibrating the CALO response to energy deposits in LYSO cubes, and together with the TRD information, it will strongly increase our understanding of the experiment energy scale. Furthermore, both systems provide independent fast trigger information which  can be employed to cross-validate the trigger efficiency. The first CALO read-out consists of WaveLength Shifting fibers (WLS) coupled to Intensified scientific CMOS (IsCMOS) cameras\cite{CaloICRC2021}\cite{HerdICRC2019}: here we present just a brief review since it will be accurately described in dedicated publications. Two WLSs optically coupled to one face of each LYSO cube are used to convert the LYSO scintillation photons into green light. Two bunches of fibers are made by grouping the ones from each cube. Each bunch is coupled to an Image Intensifier (II) that converts photons to electrons, with a different gain level for the two IIs to extend the overall dynamic range. The electrons are converted to photons again after hitting a phosphor screen and the light output of each II is finally read-out by a dedicated IsCMOS cameras.  The WLSs are also coupled to several Photo Multiplier Tubes (PMT) which provide fast trigger information regarding several CALO regions.\\
This paper describes the second read-out system: the main design, the performance of the system and the assembly of a first prototype including the double read-out are described in sections \ref{sec:PD_readout}, \ref{sec:performance} and \ref{sec:prototype}, respectively.

\section{Design of the photo-diode read-out system.}
\label{sec:PD_readout}

\begin{figure}[th]
\centering 
\includegraphics[width=.99\textwidth]{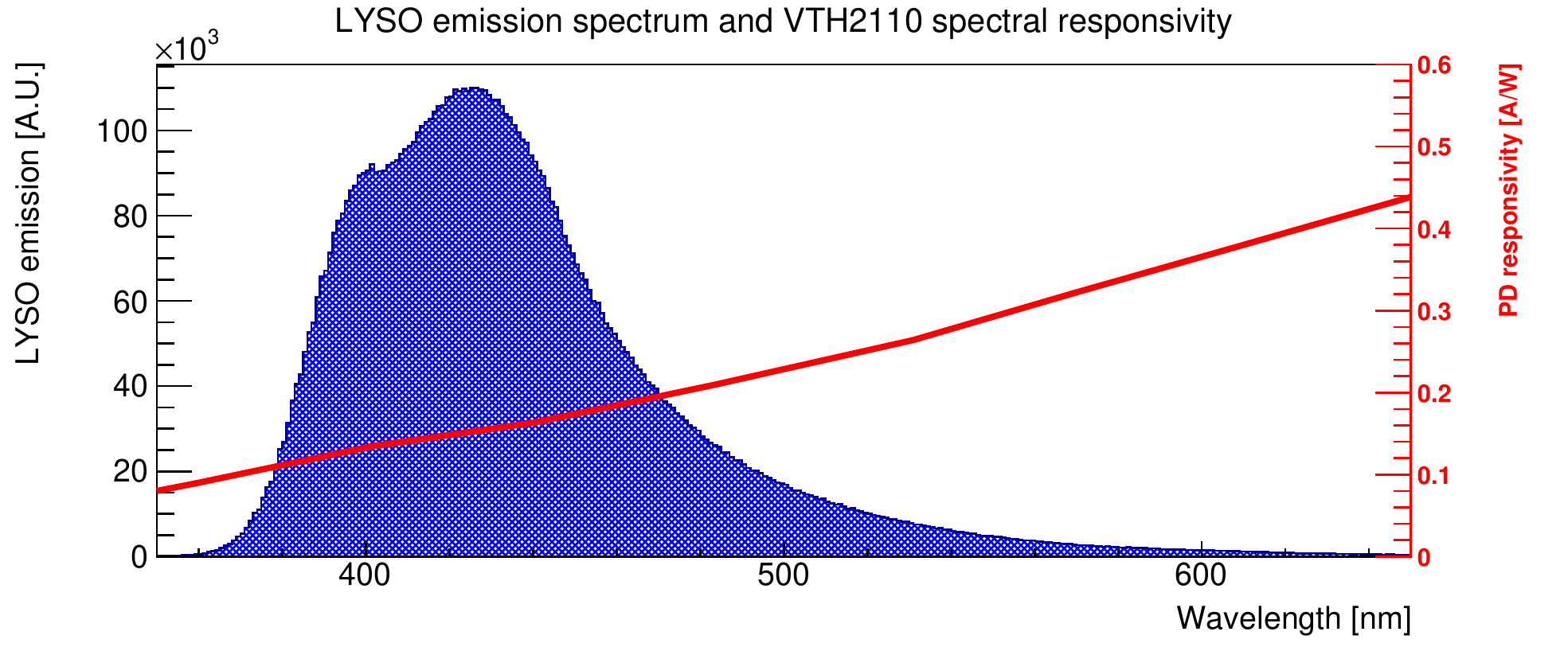}
\caption{\label{fig:LYSOspec} The LYSO emission spectrum (blue histogram) and the VTH2110 spectral responsivity (red line).}
\end{figure}

The second read-out system is based on photo-diodes (PDs), exploiting the studies carried out by the CaloCube collaboration\cite{CALOCUBE_HW}\cite{Calocube_proc3}. PDs are well suited for calorimetric space application since these sensors do not require high voltage power supplies (like typical PMTs) and accurate temperature control system (like SiPMs). A pair of PDs with different active areas are coupled to each cube: the large PD (LPD) is the Excelitas VTH2110 and the small one (SPD) is the VTP9412 whose active areas are $25\ mm^2$ and $1.6\ mm^2$, respectively. As explained in the following sections of this paper, the performance of this read-out system well meets the experiment requirements. The most relevant PD features (measured for a bias voltage of about $50\ V$) are summarized in table \ref{tab:PD}, whereas fig. \ref{fig:LYSOspec} shows the spectral responsivity of VTH2110 compared with the LYSO emission spectrum.
\begin{table} [h]
	\begin{tabular}{| c | c | c | c | c | c |}
		\hline	
		Part number & Active area & Junction cap. & Dark current & Response @ $420\ nm$ & Rise time\\ \hline	
		VTH2110 & $25\ mm^2$ &  $<30\ pF$ &  $<0.2\ nA$ & $\sim 0.15\ A/W$ & $\sim15\ ns$ \\ \hline	
		VTP9412 & $1.6\ mm^2$ &  $<6\ pF$ & $<7\ nA$ & $\sim 0.2\ A/W$ & $\sim15\ ns$ \\ \hline	
	\end{tabular}
\caption{\label{tab:PD} Main PD features measured for a bias voltage of about $50\ V$.}
\end{table}
\begin{figure}[h]
\centering 
\includegraphics[width=.99\textwidth]{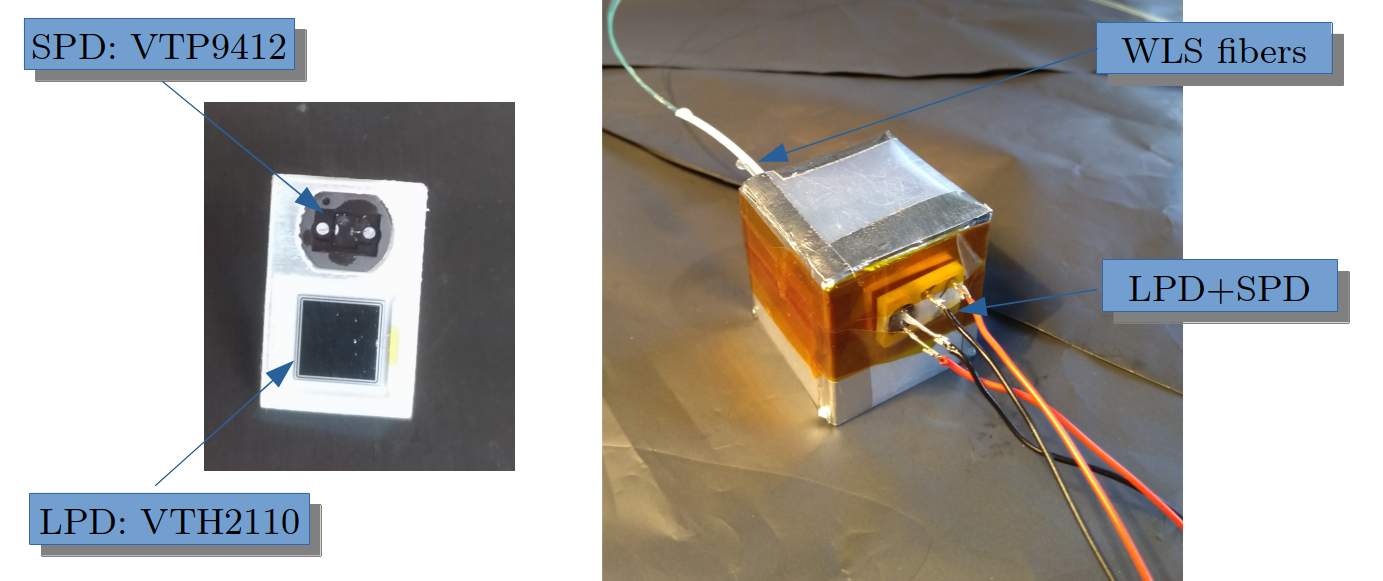}
\caption{\label{fig:PDphoto} Left picture: the large PD (VTH2110) and small PD (VTP9412) inside the preliminary package made of plastic material. Right picture: a LYSO cube wrapped with the Enhanced Specular Reflector (ESR); PDs are glued on a lateral face of the cube and WLSs are placed on the top face.}
\end{figure}\\
Due to the different active areas, the expected ratio between signals of LPD and SPD is larger than 10. In order to further increase the signal ratio, an optical filter with a transmittance of about $1.5\%$ will be added to the SPD active area: the optimization of filter transmittance value is discussed in section \ref{subsec:Overall_performance}. A monolithic, custom-designed package housing both the silicon sensor of the VTH2110 and the one of the VTP9412 will be used for assembling the flight detector: a preliminary version of the package was made in INFN Florence laboratory by incorporating the full VTH2110 and VTP9412 devices in a plastic frame and is shown in fig. \ref{fig:PDphoto}. As shown in the right panel, the LYSO crystals will be wrapped with Enhanced Specular Reflector (ESR) with a reflective power $\geq 98\%$ in the visible range. The final version of the PD assembly putting together the monolithic package, the sensors and the optical filter, is under development in collaboration with Excelitas: the current status is briefly described in section \ref{subsec:Overall_performance}. The sensors will be glued to LYSO crystals: the main candidate material for this procedure is EPOTEK-301, which is a well known optical glue suited for space applications, but other options will be evaluated, e.g. Sylgard 184 Silicone Elastomer. \\
\begin{figure}[th]
\centering 
\includegraphics[width=.99\textwidth]{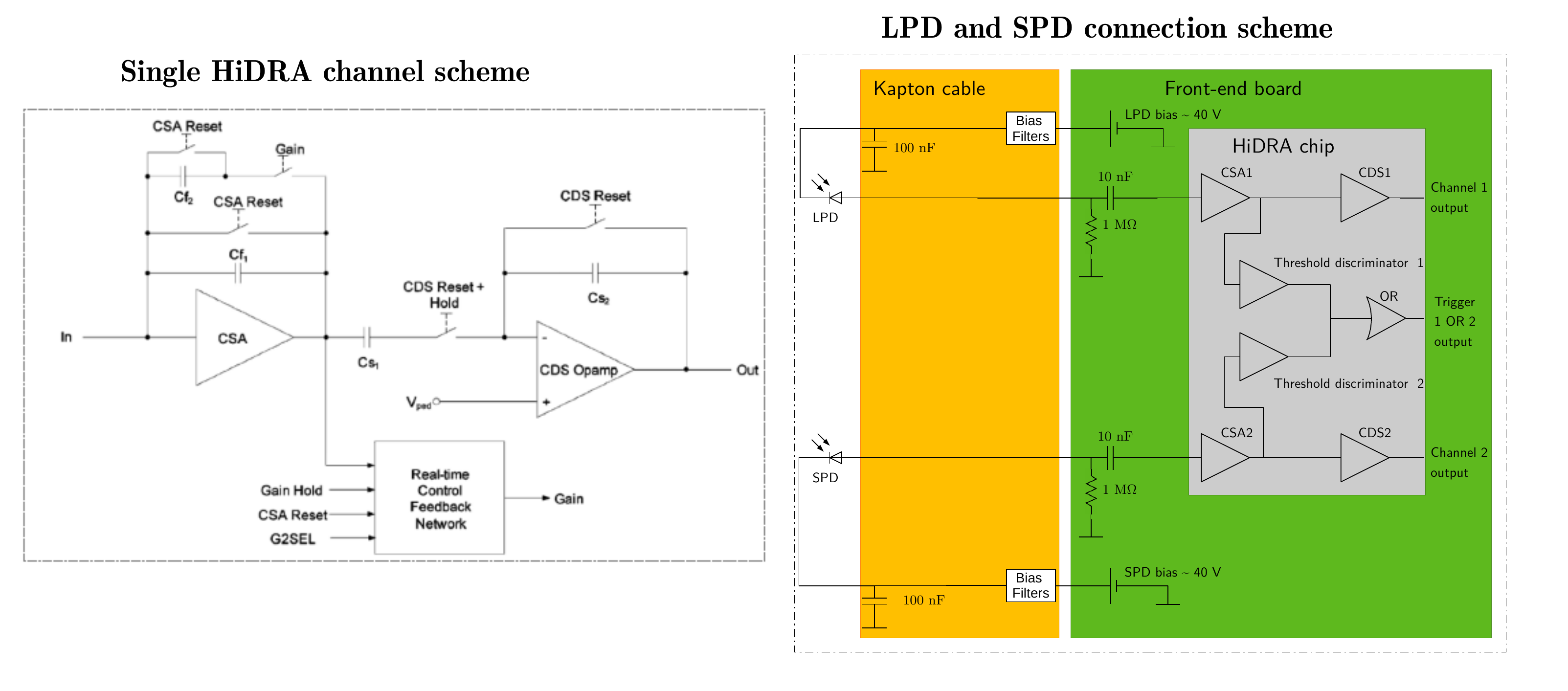}
\caption{\label{fig:ConnectionScheme} Simplified scheme of a single channel of HiDRA (left panel, here the self-trigger section is not included) and of a PD couple connections (right panel). Only the main components of the kapton cable, the FEE and two channels of HiDRA chip are included.}
\end{figure}
PDs are connected to custom front-end electronics (FEE) chips through flat kapton cables. The main component of the electronics is the HiDRA-2 chip, developed by INFN Trieste for space application. The simplified scheme of the front-end section of HiDRA chip for a single channel is shown in fig. \ref{fig:ConnectionScheme} left panel. It is an upgraded version of the CASIS chip\cite{CASIS} and HiDRA-1 to properly fit the PD system requirements. The main features of the chip are low noise (about 2500 equivalent electrons), low power consumption ($3.75\  mW/channel$), and large dynamic range (from few fC to $52.6\ pC$). It consists of a charge sensitive amplifier (CSA) with an automatic double-gain selection: when the CSA output signal is larger than a given threshold, the gain is decreased by a factor of 20. Laboratory tests show that the uncertainty of this value is below $1\%$. The output of the CSA is then connected to a correlated double sampling circuit (CDS), which is employed to sample the signal (the integration windows is few  $\mu s$), and to a fast trigger circuit. The latter comprises a common threshold generator and, for each channel, a processing chain made of a signal amplifier and a comparator. In order to decrease the power consumption and the number of chip output pins, the output of two successive comparators are ORed together and then converted to differential CMOS signals for further external processing. LPD and SPD will be connected to nearby channels, as shown by right panel of fig. \ref{fig:ConnectionScheme}: the LPD signal dominates the OR between the channel couple, thus a trigger information for each cube read-out by LPD is allowed even if the number of trigger outputs are half of the chip analogical inputs. Thanks to the granularity, this trigger system will provide fast information for a topological analysis of the shower at trigger time, enhancing the system capabilities of selecting specific particle species and energies at the trigger level. Each HiDRA-2 chip has 16 input channels whose analog output is multiplexed to a single output pin. A detailed description of the front-end electronics calibration procedure can be found in \cite{CALOCUBE_ELECTRON}.\\ 
\begin{figure}[th]
	\centering 
	\includegraphics[width=.9\textwidth]{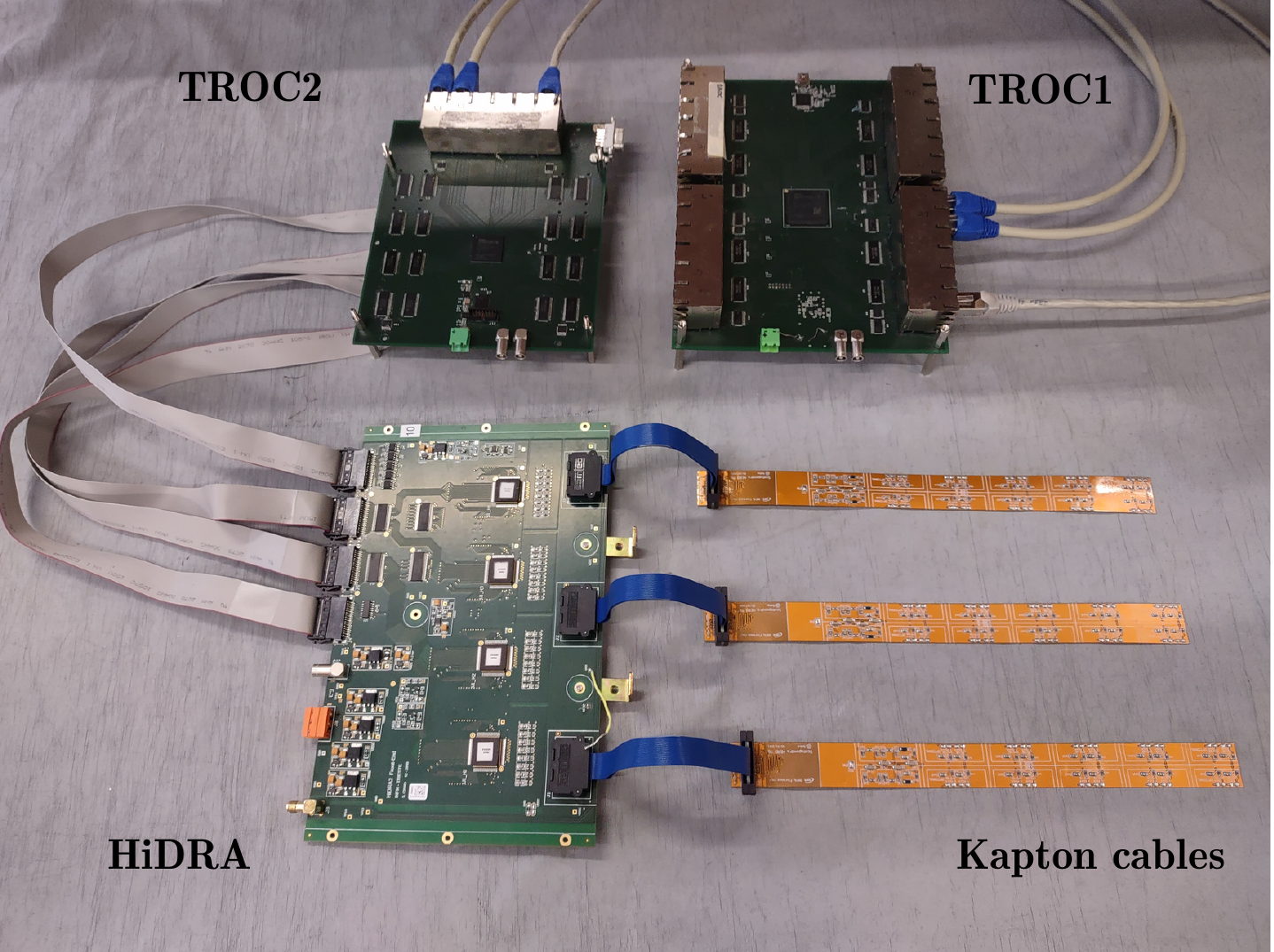}
	\caption{\label{fig:HIDRAPicture} A picture of the 2020-2021 version of DAQ boards (TROC1 and TROC2), FEE board (HiDRA) and kapton cables.}
\end{figure} \begin{figure}[th]
	\centering 
	\includegraphics[width=.99\textwidth]{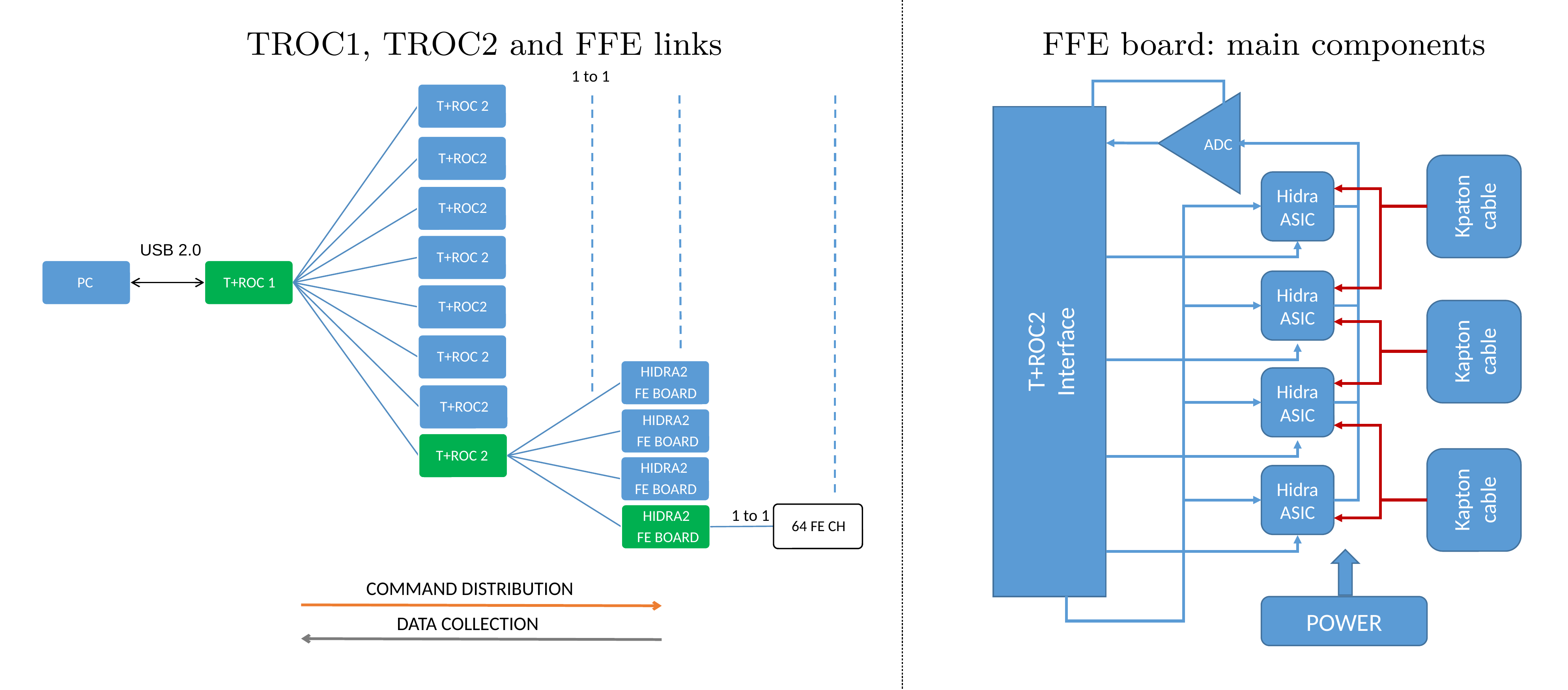}
	\caption{\label{fig:TRCOscheme} Simplified scheme of connection between DAQ and FEE boards (left panel) and main components of a FEE board (right panel).}
\end{figure}The FEE is driven by a data acquisition system (DAQ) developed by CIEMAT-Madrid. The main goals of this system are: to merge all the information coming from single cubes; to configure each part of the system to a given acquisition mode; to monitor the proper functioning of each part of the system; to decide if the signal of the complete calorimeter could be significant and transmit the decision fast enough to the general trigger system.\\
A first version of the DAQ and FEE showed in fig. \ref{fig:HIDRAPicture} was developed and employed for the tests described in this paper. The main components of a FEE board are four HiDRA-2 chips and a single ADC, which converts all the 64 channel signals: the connections between these components are shown in fig. \ref{fig:TRCOscheme} right panel. Due to mechanical constraints of the current version of CALO prototype, three kapton cables are employed in order to connect up to 60 PDs to the four chips, thus 4 channels are not properly connected. This configuration will be updated for the next version of the prototype (which will be assembled in 2023) and PD system in order to optimize the number of used HiDRA channels. Furthermore, the FEE is connected to the kapton cables by using short Samtec flat (blue) cables: these are employed at this stage of the project to simplify the assembly of different prototypes, where the alignment of kapton cables and FEE board connectors is not guaranteed. For the flight detector these additional cables will be eventually removed or replaced, depending on the final design of the FEE+DAQ system.\\
The DAQ consists of a master board (named TROC1) and eight slave boards (TROC2): the main links between DAQ and FEE boards are shown in the left panel of fig. \ref{fig:TRCOscheme}. The TROC1 is in charge of building the event with the information received from the TROC2 boards, and distributes the  configuration commands. The main goals of the TROC2 are to generate the digital signal sequence needed to drive HiDRA-2 and ADC, and to transmit the ADC and self-trigger data to the TROC1. Each TROC2 can be connected to 4 HiDRA boards, therefore the maximum number of channels read-out with this system version is about 2000. 

\section{Performance of the HERD PD system.}
\label{sec:performance}

In this section the main features of the HERD PD read-out system obtained with several measurements are discussed: a few comparisons with the previous results obtained by the CaloCube collaboration are also shown.

\subsection{PD signals.} 
\label{subsec:PDsignal}

\begin{figure}[h]
	\centering 
	\includegraphics[width=.99\textwidth]{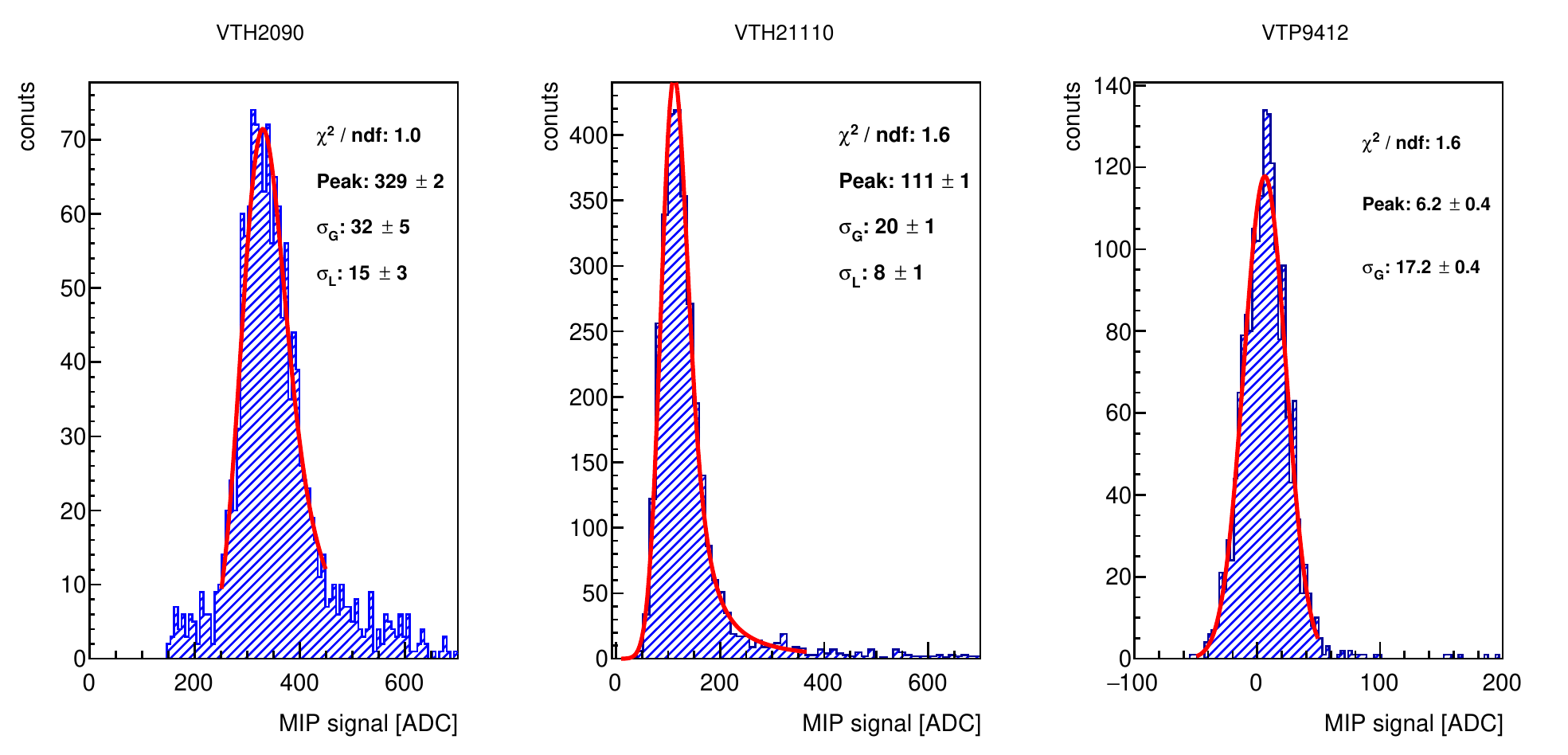}
	\caption{\label{fig:PDMIPs} MIP signal obtained with three PDs glued to LYSO crystals and read-out with the HiDRA FEE. Left panel: large PD (VTH2090) used for the CaloCube prototype. Central and right panels: large and small PDs which will be employed for HERD. Starting from the left, the first two histograms are fitted with convoluted Landau and Gaussian functions, while the last one is fitted with a Gaussian function.}
\end{figure}

\begin{figure}[t]
\centering 
\includegraphics[width=.99\textwidth]{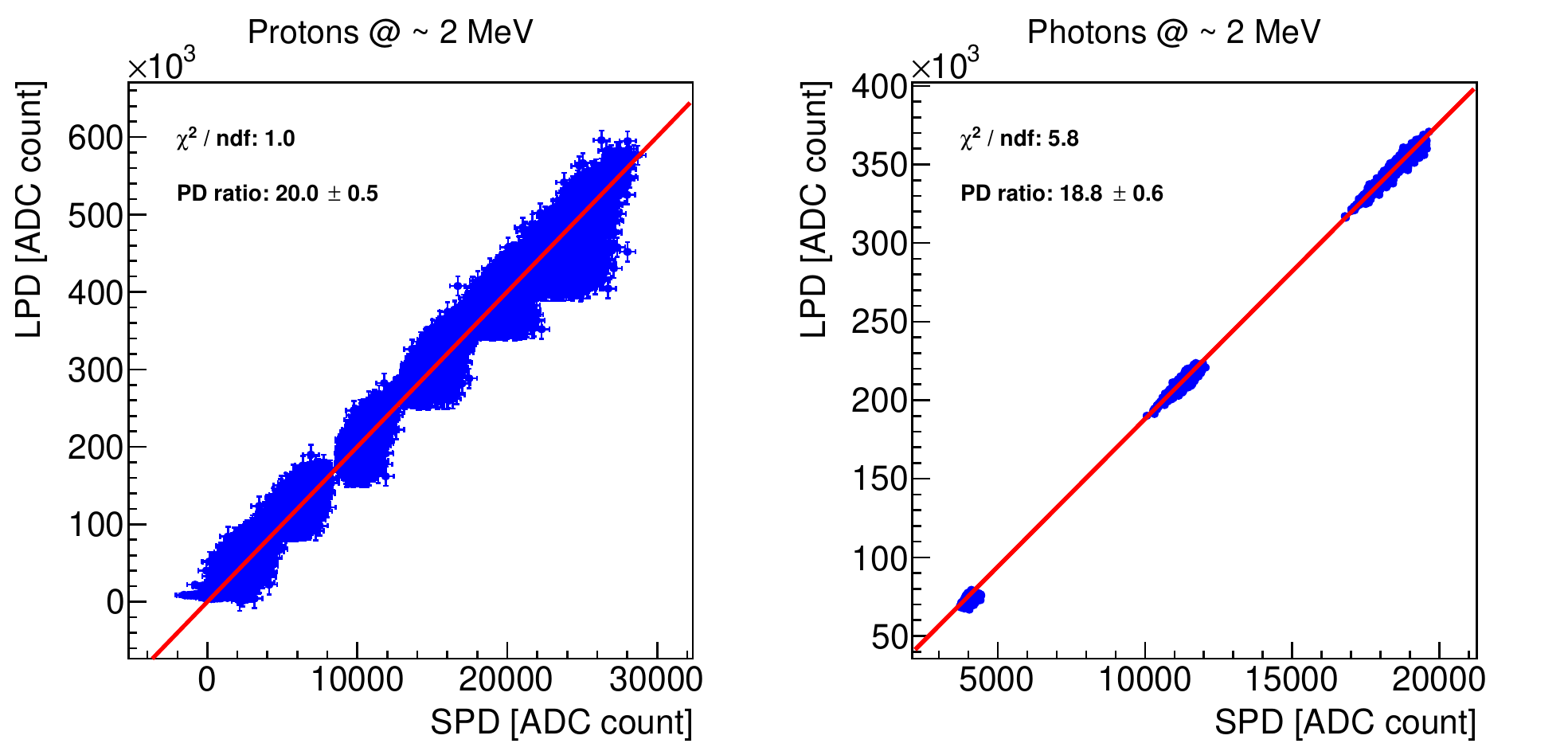}
\caption{\label{fig:PDratio} Left panel: LPD signal as a function of the SPD one, for each events acquired with low energy protons, with the red line representing the fitted function. Right panel: same graph obtained with low energy photon data and a different LYSO+PDs sample.}
\end{figure}

Since the CALO channels will be calibrated using non-interacting protons and nuclei, the minimum scintillation light which must be detected by the read-out system is the one corresponding to a Minimum Ionizing Particle (MIP). By taking into account the main crystal characteristics summarized inside table \ref{tab:LYSO}, the energy deposit by a MIP in $3\ cm$ of LYSO is about $30\ MeV$, which corresponds to  $\sim9\cdot 10^5$ photons distributed around the maximum of the LYSO emission spectrum. Fig. \ref{fig:PDMIPs} features the MIP signals obtained by vertical ground muons which hit LYSO cubes: the signal distributions of three PDs are shown. Fig. \ref{fig:PDMIPs} left panel features the signal obtained with the large PD (VTH2090) employed for the CaloCube prototypes: it has an active area of $92.16\ mm^2$ and very similar quantum efficiency with respect to the PDs used in HERD. By using the HiDRA-2 FEE the MIP peak corresponds to about $330\ ADC\ count$. Since the production of VTH2090 is discontinued, it cannot be used for HERD. The distributions of typical MIP signals obtained with the new LPD (VTH2110) and SPD (VTP9412) are shown in the central and right panels of fig. \ref{fig:PDMIPs}. The LPD signal is smaller than the one obtained with VTH2090: it is mainly due to the different active area.
\\
As expected, several parameters (e.g. crystals light output, PD and LYSO coupling, wrapping reflection efficiency,...) will affect the response of the system: these parameters could vary among different crystals and PDs. Thus, to properly estimate the MIP peak value, several tests with different PDs and LYSO crystals were performed. Finally, we computed the mean and the standard deviation (SD) of the distribution of MIP peak values obtained with both LPD and SPD. In this paper, these values are considered as the best estimation of mean value and dispersion of the MIP peak. By using the described analysis procedure, the mean and dispersion of MIP peak obtained with LPD are $110\ ADC\ count$ and $10\ ADC\ count$, while the mean and dispersion obtained with the SPD are $6\ ADC\ count$ and $2\ ADC\ count$, respectively. The large SPD dispersion is mainly due to a systematic error related to the variation of the pedestal during long measurements (about one week for each sample of LYSO+PDs), which affects the SPD due to the small S/N ratio, while it is negligible for LPD. By taking into account the mean value of the distributions obtained with these measurements, the ratio between LPD and SPD signals (LPD/SPD) is about 19, but energy deposits inside crystals larger than few MIPs are required to properly confirm this rough estimation. Thus, we tested two samples of LYSO cubes with preliminary sensor packages by employing low energy and high intensity particle beams. The first one was provided by the Labec facility\cite{Labec}: the beam consists of low energy protons with a kinetic energy of $2\ MeV$. By properly adjusting the beam configurations, it is possible to investigate LPD/SPD corresponding to different energy regions. The maximum available number of particles deposits an energy inside a LYSO crystal which could saturate the PD system. The crystals and PDs were contained in a vacuum chamber. Dedicated custom cables were developed for the particular configuration. The correlation between a LPD and SPD signals coupled to the same LYSO crystal is shown in the left panel of fig. \ref{fig:PDratio}: the graph includes all data acquired with different beam configurations.
 The error bars are equal to the noise of the corresponding channels estimated with the SD of the PD pedestals. The spread and asymmetric distribution of the points is due to large electromagnetic interference caused by the vacuum chamber and accelerator components. The data are then fitted with a line (red) and the angular coefficient represents the LPD/SPD ratio, which is $20.0 \pm 0.5$. To confirm the result, LYSO crystals were also tested using low-energy high-intensity photon beams provided by an ELEKTA VERSA HD linear accelerator located at the University Hospital of Florence. The noise which affected the system was smaller than the one observed at Labec and it was similar to the one measured during test with atmospheric muons. The beam intensity was not arbitrarily configurable, thus only small sections of the PD system signal range were investigated with this method. An example of the LPD/SPD ratio obtained with the photon beam is shown in the right panel of fig. \ref{fig:PDratio}: here 3 different beam intensities were employed. The average LPD/SPD ratio estimated by using the 2 methods (low energy protons and photons) for few samples of LYSO crystals is $19.5$, while the ratio SD (i.e. the dispersion) is $1.5$.

\subsection{Front-end electronics: noise and trigger efficiency.} 
\label{subsec:NoiseTrigger}
The characterization of FEE employed by CaloCube is summarized in \cite{CALOCUBE_HW}, while the performance measured with high energy particle beams are discussed in \cite{CALOCUBE_ELECTRON}\cite{Calocube_proc4}\cite{Calocube_proc5}. In view of the HERD experiment, several updates of the electronics were carried out. Two main items are discussed in this section: the noise value, which was decreased, mainly thanks to the optimized kapton cable design, and the new HiDRA-2 self-trigger circuitry performance.\\
\paragraph{Noise.} 
\begin{figure}[h]
	\centering 
	\includegraphics[width=.99\textwidth]{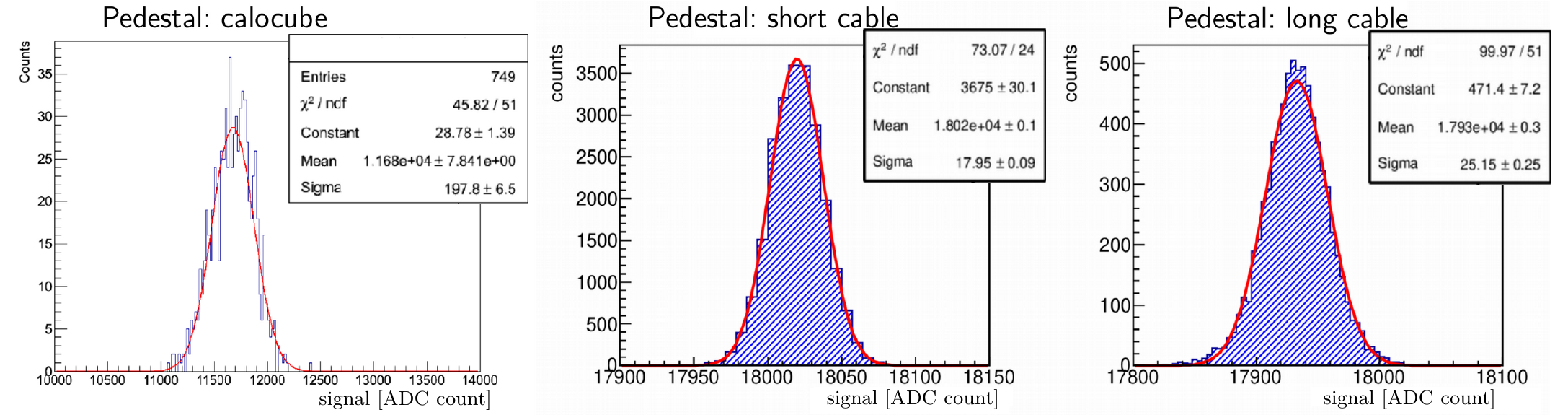}
	\caption{\label{fig:Noise} Pedestal distributions fitted by Gaussian functions. Left panel shows a typical pedestal obtained by the CaloCube project\cite{CALOCUBE_HW}. The central and right panels show the typical pedestals obtained with the HERD short and long kapton cables, respectively.  }
\end{figure} 
\begin{figure}[h]
	\centering 
	\includegraphics[width=.99\textwidth]{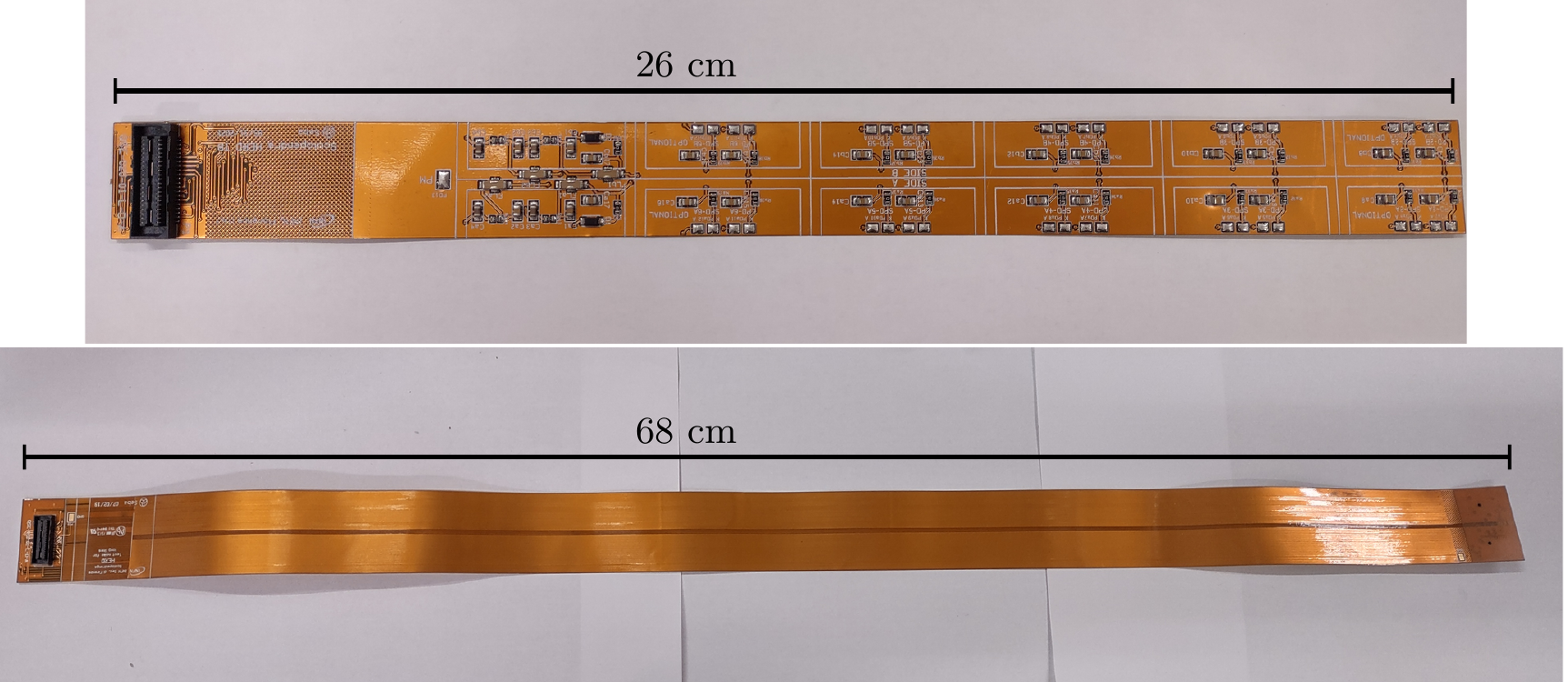}
	\caption{\label{fig:kaptonCable} Top panel: a short kapton cable designed for the CALO prototype. Bottom panel: a long kapton cable employed to emulate the CALO flight configuration. }
\end{figure}The noise observed with the FEE developed for CaloCube was dominated by the common-noise (CN), which was mainly due to the coupling between different traces. A typical value of the noise was about $180\ ADC\ count$, see left panel of fig. \ref{fig:Noise}: here the noise is estimated as the SD of the pedestals, i.e. the distribution of output signals obtained without input signals. The intrinsic noise, which was estimated with a CN subtraction algorithm, was $~20\ ADC\ count$. A picture of the new kapton cable developed for HERD prototype (see section \ref{sec:prototype}) is shown in top panel of fig. \ref{fig:kaptonCable}. The design was driven by two main goals:
\begin{itemize}
	\item decreasing the number of intersections between different copper traces, which strongly affects the CN,
	\item increasing the ground shielding between signal connections: it further decreases the traces capacitance and the impact of external electromagnetic interference.
\end{itemize} 
An improvement of the performance is obtained: by using the same PDs, the total noise (i.e. CN + intrinsic noise) measured with the new cable is $\sim18\ ADC\ count$, which is about a factor $6$ smaller than typical MIP peak values obtained with the VTH21110. A pedestal distribution is shown in the central panel of fig. \ref{fig:Noise}. This short cable is capable of reading out 20 PDs and it is $26\ cm$ long. The kapton cable for the flight detector will be longer, since it will be connected to about 42 PDs. In order to estimate the expected noise of the flight model, a preliminary version of the longer kapton cable was developed (bottom panel of fig. \ref{fig:kaptonCable}). It is mainly employed to increase the copper trace length of about $65\ cm$, which strongly affects the amplifier equivalent input capacitance. With this new configuration, the total noise mean and SD obtained by testing hundreds of different channels are $27.5\ ADC\ count$ and $2.5\ ADC\ count$, respectively. A typical pedestal distribution is shown in the right panel of fig. \ref{fig:Noise}. The performance of FEE is further improved if a CN subtraction algorithm is applied. Each kapton cable includes two diodes (blind with respect to the light) which are designed to sample a large fraction of the CN event-by-event. By using this information, the typical noise achieved with the CN subtraction is $\sim22.5\ ADC\ count$. Since the performance of self-trigger circuit can not benefit of CN subtraction, as a conservative estimation, in the following sections of this paper the total noise is used for the estimation of S/N ratio and dynamic range.\\
The accurate design of the cable also features a very small cross-talk between nearby traces. This was measured with two methods: the first one exploited a high intensity laser to inject signals in a PD connected to a single channel and check the response of near-by channels. The second method consisted in isolating the contribution of the kapton cable: by properly modifying it, pulses of different shapes and intensity were injected inside a trace, and the signal obtained in nearby traces was acquired. With these two methods, the obtained cross-talk was smaller than $0.1\%$ with respect to the injected signal, which confirms the excellent isolation of traces.\\
A long kapton cable was also employed during test with ground muons: the results are compatible with the ones discussed in section \ref{subsec:PDsignal}, which are obtained with very short cables. Thus, as expected, the system response is not strongly affected by the long copper traces.

\paragraph{Self-trigger.} 
\begin{figure}[ht]
	\centering 
	\includegraphics[width=.99\textwidth]{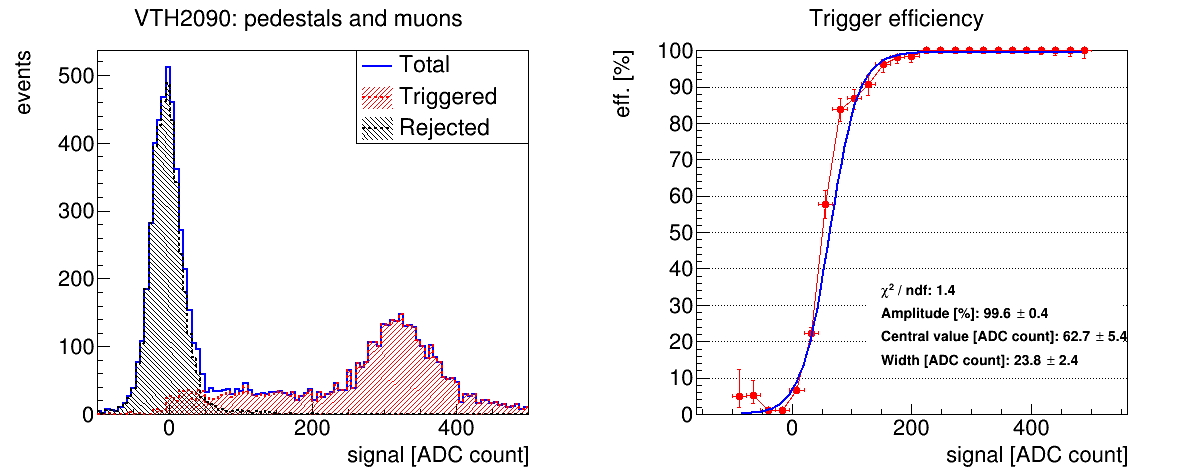}
	\caption{\label{fig:Trigger} The left panel shows the signal distribution obtained with a LYSO crystal read-out by a VTH2090 PD: blue histogram includes all the acquired events, red (black) histogram includes events which are accepted (rejected) by the HiDRA self-trigger circuit. The right panel shows the trigger efficiency as a function of the signal amplitude: the graph is fitted using a logistic function.   }
\end{figure}A new feature developed for the HERD experiment is the self-trigger circuit included inside the HiDRA-2 chip. Four values of the trigger threshold, equal to an internally generated threshold voltage multiplied by 1, 1.5, 2 and 2.5 are available by properly setting the corresponding chip digital inputs. These different thresholds could be employed during in-orbit operations in order to optimize the trigger configurations for specific runs, e.g. the high threshold is useful to select non-interacting helium nuclei, while rejecting most of non-interacting protons. MIP signals obtained with atmospheric muons were employed to check the trigger efficiency. Fig. \ref{fig:Trigger} shows the MIP signal distribution (left) obtained with VTH2090 PD coupled to a LYSO crystal and the corresponding trigger efficiency (right), obtained by selecting the lower trigger threshold. The graph is fitted using a logistic function. The obtained trigger thresholds is $63\pm5\ ADC\ count$, the function steepness is about $24\ ADC\ count$ and the fraction of events triggered by the noise is lower than $5\%$. By taking into account the fit errors, both mean and steepness are consistent among measurements obtained with different HiDRA channels and chips. By considering the MIP value obtained with LPD, the minimum trigger threshold is about half of MIP peak. This is a reasonable value for main goals of the PD trigger system, which are to select low-energy shower and not-interacting hadrons. A preliminary trigger scheme, which includes the PD system, is described in \cite{Trigger}.

\subsection{Signal-to-noise ratio and saturation level.} 
\label{subsec:Overall_performance}

By taking into account the results described in the previous paragraphs, the expected signal/noise (S/N) ratio for the MIP peak obtained with the current flight configurations of PD and FEE, is about $4$. This allows a proper calibration of LYSO cubes with signals provided by CR protons which do not start a shower inside the crystals. By using the preliminary monolithic package, the saturation of LPD and SPD are  $\sim 190\ GeV$ and $\sim 3.5\ TeV$, respectively. As described in section \ref{sec:PD_readout}, the new package design for the flight detector will include an optical filter on the SPD. The filter transmittance value has to be carefully chosen since it strongly affects both the saturation level and the overlapping region of LPD and SPD signals. The latter should be large enough to allow proper calibrations of SPD as a function of LPD signal using high energy showers. For instance, a small transmittance allows you to reach a high saturation level but a small LPD-SPD overlap region. The selected value, which is $1.5\%$, is a compromise between these two opposite requirements. For the SPD, a very high saturation level of $250 \pm 40\ TeV$ is achieved. The corresponding dynamic range of the system is about $3\cdot 10^7$, which is larger than the one obtained with previous space calorimeter read-out systems, e.g. CALET features a dynamic range of $10^6$\cite{CALET_calibration} while Fermi-LAT and DAMPE have saturation levels of single calorimeter elements equal to about $70\ GeV$\cite{FERMI_Calo} and $10\ TeV$\cite{DAMPE_Calo}, respectively. The filter transmittance choice guarantees a large LPD-SPD overlap as well, since the S/N ratio of SPD is larger than $15$ for a signal which saturates the LPD channel. The large dynamic range and the capability of an accurate and complete system calibration are the main goals achieved with the work discussed in this article: both results will have a sizable positive impact on the performance of the HERD experiment. 
\begin{figure}[t]
	\centering 
	\includegraphics[width=.99\textwidth]{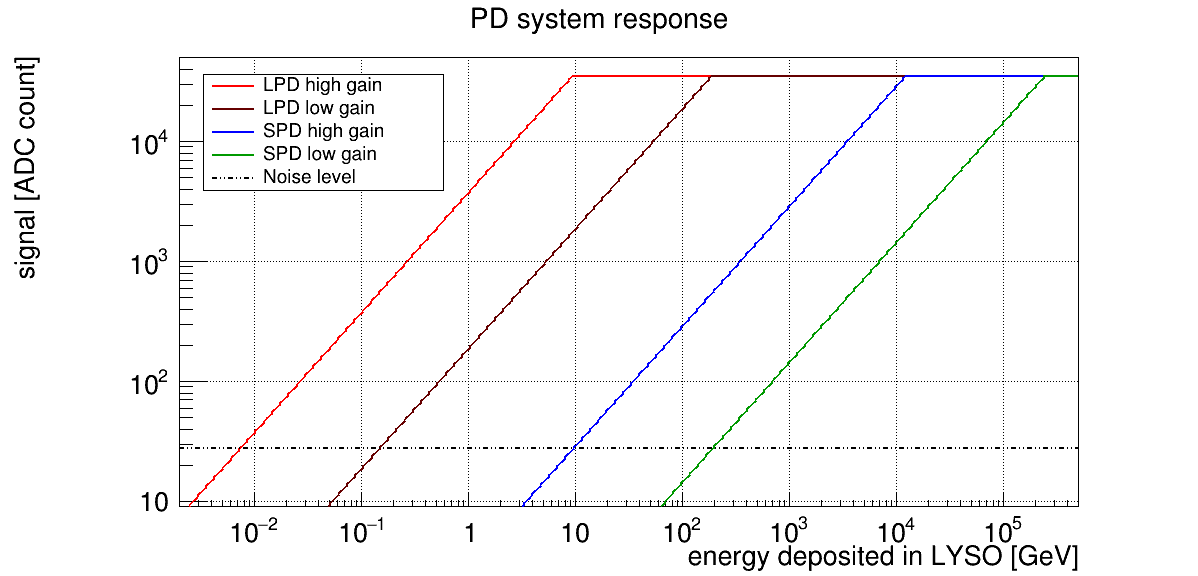}
	\caption{\label{fig:ranges} Expected output signal of PD system as a function of the energy deposit in LYSO crystal. The response of LPD and SPD, acquired both in high and low gain, is shown. The noise level is also include.}
\end{figure}
Fig. \ref{fig:ranges} shows the expected PD read-out response as a function of the energy deposited inside a LYSO crystal, while the table \ref{tab:final} summarized the main features of the system.
\begin{table} [h]
\begin{tabular}{| c | c | c | c | c | c| c|}
	\hline	
	LPD MIP & Noise & MIP S/N ratio & Filter & LPD/SPD & SPD S/N & Saturation \\ \hline	
	$110\pm10\ ADC$ & $<30\ ADC$&  $>3.5\ ADC$ & $1.5\%$ & $1300 \pm 100$ & $>15 $ & $250 \pm 40\ TeV$ \\ \hline	
\end{tabular}
\caption{\label{tab:final} Main features of the PD read-out system.}
\end{table}
\begin{figure}[h]
\centering 
\includegraphics[width=.99\textwidth]{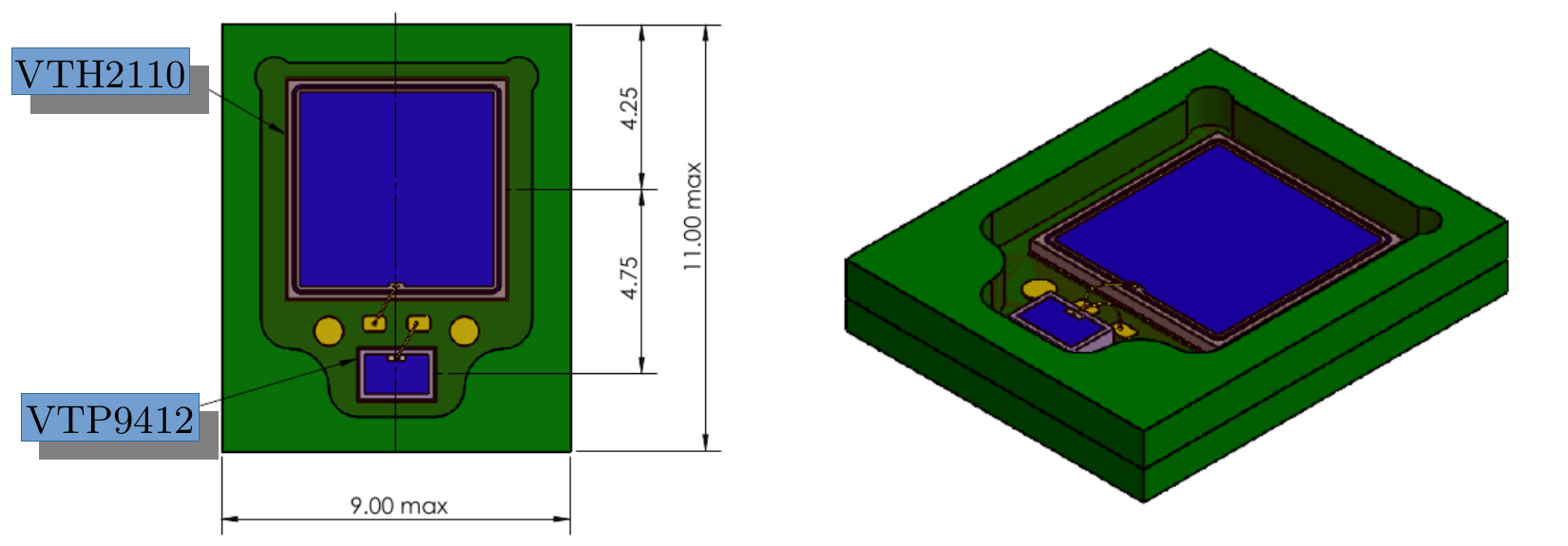}
\caption{\label{fig:NewPackage} Two images of the new monolithic package which includes LPD, SPD and the optical filter. Courtesy of Excelitas.}
\end{figure}\\
Since all the system parameters are appropriate with respect to the experiment requirements, the production of the new monolithic package for PDs will start during the first months of 2022 in collaboration with Excelitas. Two images of the design of the new product are shown in fig. \ref{fig:NewPackage}. It includes the active area of the two PDs (VTH2110 and VTP9412) on the top face, four pads connected to the PD anodes and cathodes arranged on the back face (the pads are not shown in the images), an optical filter which completely covers the SPD active area, and an optical resin on the top face. The carrier is made of FR4 and the optical filter transmittance is $(1.5 \pm 0.1)\%$ for $[410;450]\ nm$ light, which corresponds to the LYSO peak emission. Finally the selected optic resin allows a transmittance larger than $95\%$ for the same wavelength range. 

\section{PD system for the CALO prototype.}
\label{sec:prototype}
\begin{figure}[h]
	\centering 
	\includegraphics[width=.95\textwidth]{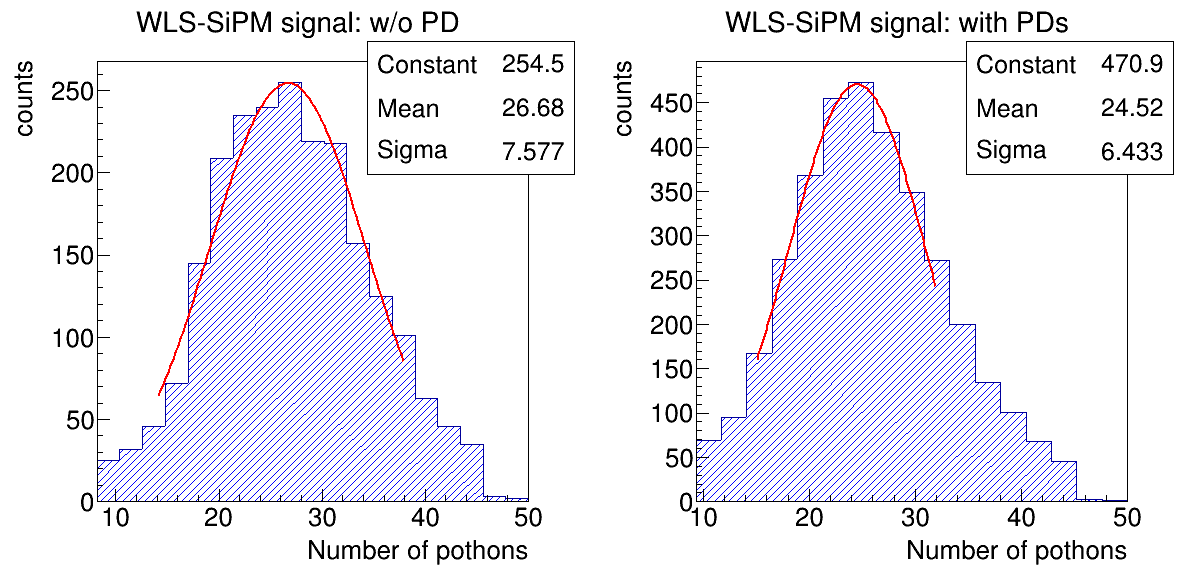}
	\caption{\label{fig:PDvsWLS} Distribution of MIP signals obtained with crystals read-out with a WLS fiber coupled to a SiPM. Left plot: PD are not glued to the cube. Right plot: the preliminary monolithic package was glued on the cube. The distribution peaks are fitted by Gaussian functions.}
\end{figure}
In order to test the CALO performance with high energy beams, a prototype made of $5 \times 5 \times 21$ LYSO crystals was assembled during the summer of 2021. This calorimeter consists of 7 horizontal layers, but only 5 of them are filled with $5 \times 21$ cubes each. All the crystals are equipped with WLSs read-out by IsCMOS. The PD system was also included inside this prototype in order to perform a first test of this unique double read-out system with high energy particles. Since PD active and passive areas absorb a fraction of the light emitted by the LYSO, WLS signals are expected to decrease when PDs are glued to the cubes. Tests done in parallel at INFN-Firenze and IHEP-Beijing laboratories show a typical fiber signal decrease $\lessapprox 10\%$, which does not strongly affect the performance of WLS+IsCMOS system. An example of the test results is shown in fig. \ref{fig:PDvsWLS}.
\begin{figure}[t]
\centering 
\includegraphics[width=.4\textwidth]{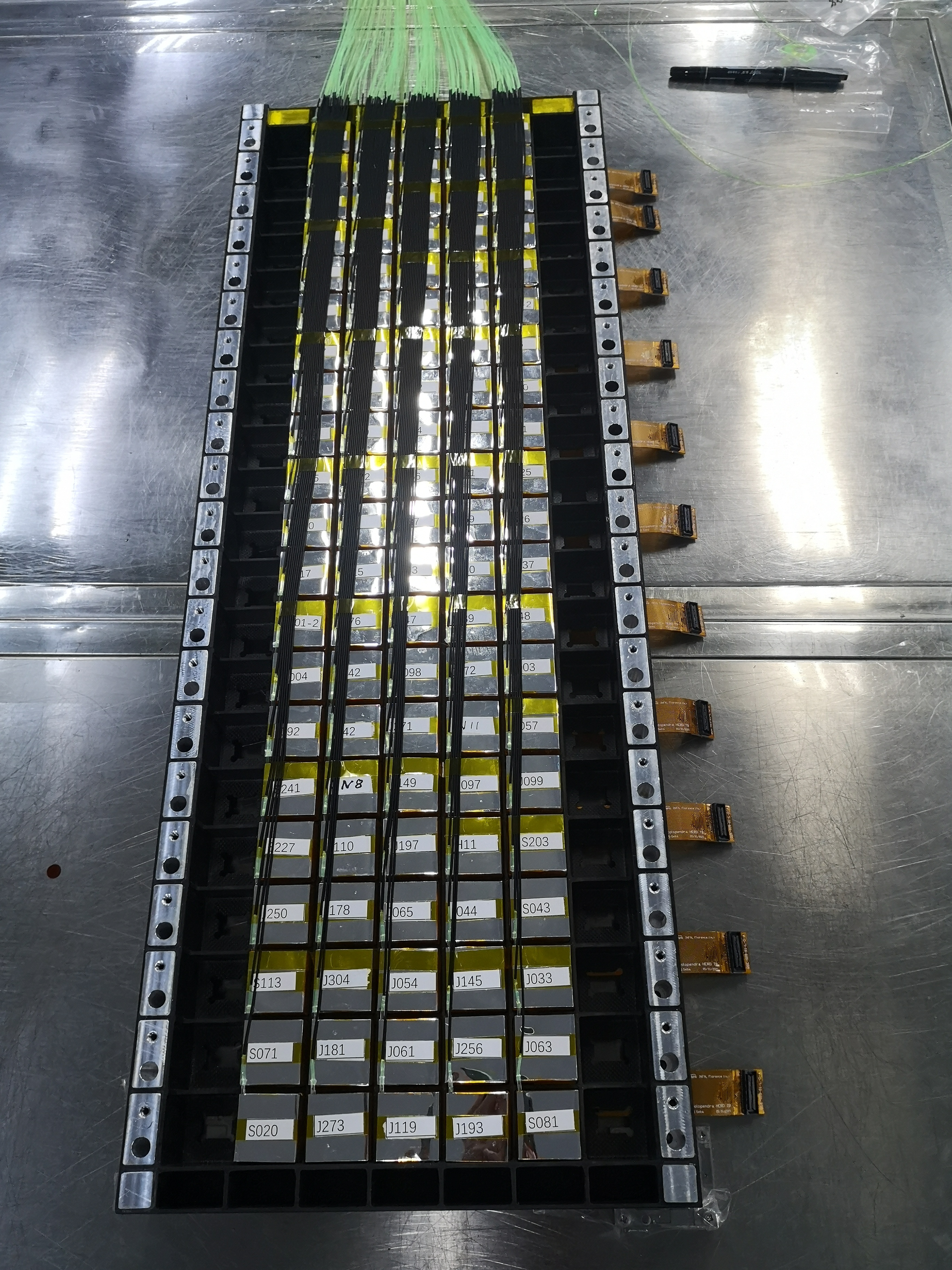}
\qquad
\includegraphics[width=.4\textwidth]{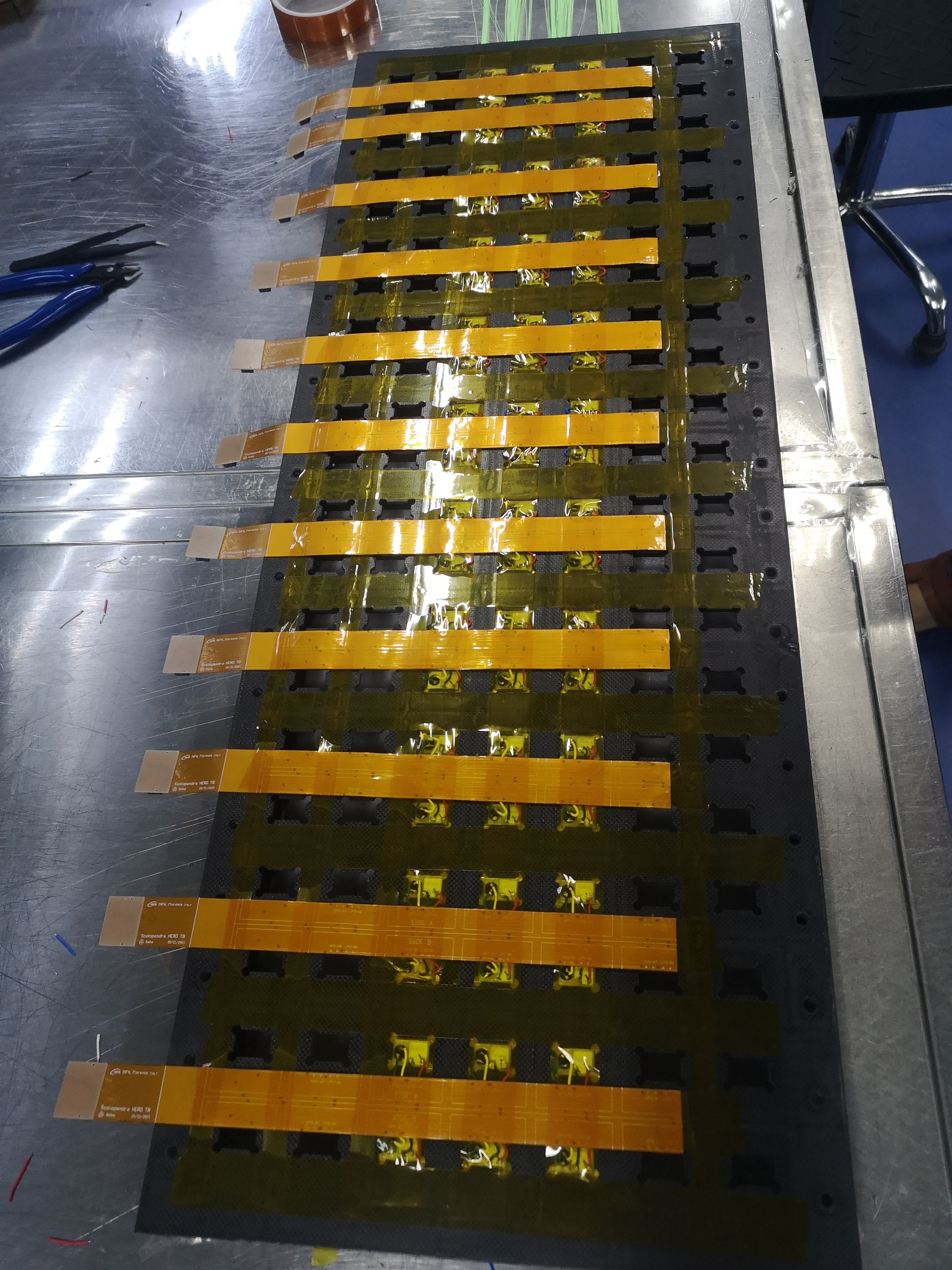}
\caption{\label{fig:Layer} Front and back view of the prototype layer which includes PDs and kapton cables.}
\end{figure}
\begin{figure}[h]
\centering 
\includegraphics[width=.99\textwidth]{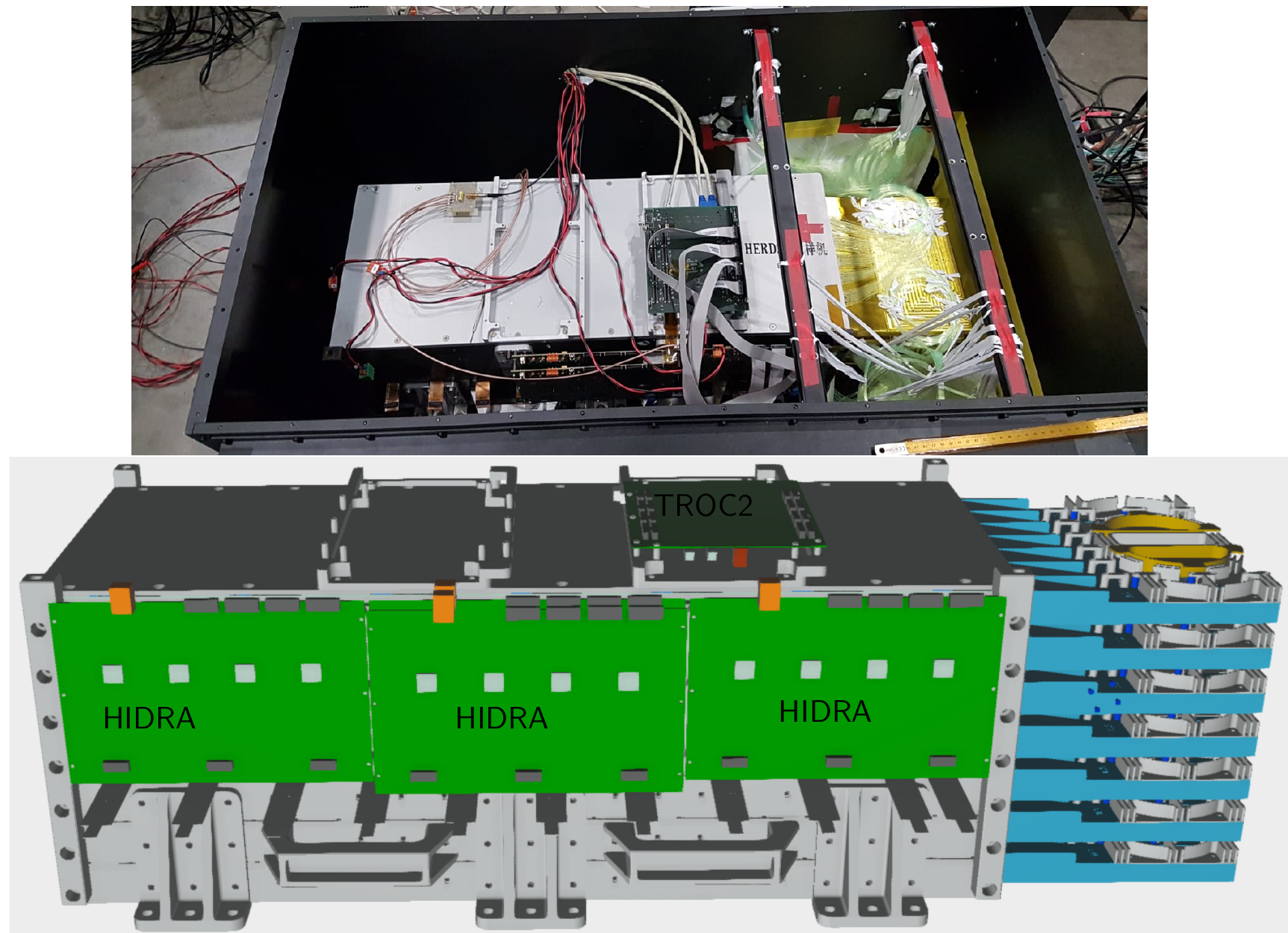}
\caption{\label{fig:PrototypeImages} The 3D design and a photograph of the CALO prototype. The latter was taken during the prototype assembly; at that time only 3 HiDRA board were presents.   }
\end{figure} A WLS fiber attached to a crystals top face is coupled with a SiPM (Hamamatsu S12571-010) and MIP signals are acquired before and after the application of a monolithic package, which included both PDs, on the bottom face. By fitting the distribution peaks, we estimate a WLS signal attenuation of $\simeq 8.3\%$ after PDs addition.\\
Due to the limited amount of LPD samples, only $3 \times 21$  cubes of the prototype are read-out with PDs. Those crystals are placed on the 3 central rows of the central horizontal layer. This configuration allows a good energy resolution for electrons (the total depth of LYSO crystals is about $55\ X_0$), and a good sampling of the longitudinal and horizontal lateral shower profile. Two pictures of the layer are shown in fig. \ref{fig:Layer}.  PDs are connected to 4 FEE boards through kapton cables (top image of fig. \ref{fig:kaptonCable}) which are perpendicular with respect to the WLSs. FEE boards are placed on a lateral side of the prototype, while a TROC2 board is placed on the top, as shown in fig. \ref{fig:PrototypeImages}. The prototype is contained in a large metal box, which is employed as an additional shield against the electromagnetic interference: the TROC1 board is placed outside the box.\\
The prototype was tested with high energy beams at the CERN Super Proton Synchrotron (SPS) in October 2021 and the data analysis is ongoing: the results of SPS beam test will be discussed in forthcoming papers. Preliminary results show good agreement with the performances discussed in this paper, e.g. MIP peak with LPD, SPD/LPD ratio, FEE noise.

\section{Conclusion.}
\label{sec:conclusion}

In this paper, we presented the design of the HERD CALO PD read-out system and several tests performed during 2020 and 2021. The results show that the system meets the experiment main requirements. A signal-to-noise ratio of about $4$ for vertical MIP traversing the LYSO cube guarantees an accurate calibration of the large photo-diode, while the overlapping region between the two sensors allows the calibration of the small photo-diode by using high energy shower signals. By taking into account similar calorimeter experiments, an unique dynamic range, of about $3\cdot 10^7$, is achieved. This implies that energy deposits smaller than about $250\ TeV$ will not saturate the read-out of a single LYSO cube. This feature will allow the first direct detection of multi $PeV$ protons showers without using software correction for crystals saturation, which usually have a negative impact on the uncertainty of the energy estimation. The new self-trigger circuit of the front-end electronics is also validated: it will be mainly used for the topological trigger of low energy electro-magnetic showers and not-interacting particles. Thank to these studies, the production of the final version of the sensors will start in 2022. A hundred of samples should be available at the end of 2022. It is expected to include the new package-PDs inside the CALO prototype for new tests with high energy particles in 2023.\\

\end{document}